\documentclass{emulateapj}

\newcommand{\msunyr}{\ensuremath{\mathit{M}_{\odot}{\rm yr}^{-1}}}   
\newcommand{\kms}{\ensuremath{{\rm km\,s^{-1}}}}                   
\newcommand{\msun}{\ensuremath{\mathit{M}_{\odot}}}   
\newcommand{\lsun}{\ensuremath{\mathit{L}_{\odot}}}                  
\newcommand{\rsun}{\ensuremath{\mathit{R}_{\odot}}}                  

\newcommand{\lstar}{\ensuremath{\mathit{L}_{\star}}}                 
\newcommand{\mdot}{\ensuremath{\dot{M}}}                             
\newcommand{\rstar}{\ensuremath{\mathit{R}_{\star}}}                 
\newcommand{\teff}{\ensuremath{\mathit{T}_{\rm eff}}}                
\newcommand{\reff}{\ensuremath{\mathit{R}_{\rm phot}}}                
\newcommand{\vinf}{\ensuremath{v_{\infty}}}                          
\newcommand{\vesc}{\ensuremath{v_{\rm esc}}}                         
\newcommand{\tstar}{\ensuremath{\mathit{T}_{\star}}}                 
\newcommand{\K}{\ensuremath{\mathrm{K}}}                         

\newcommand{\vrot}{\ensuremath{v_{\rm rot}}}                         
\newcommand{\vcrit}{\ensuremath{v_{\rm crit}}}                         
\newcommand{\vphot}{\ensuremath{v_{\rm phot}}}                         
\newcommand{\vs}{\ensuremath{v_{\rm s}}}                         
\newcommand{\taulyc}{\ensuremath{\tau_{\mathrm{Lyc}}}}                 

\shorttitle{On the nature of AG Car II. Evolution close to the Eddington and bistability limits}
\shortauthors{Groh, Hillier \& Damineli}

\begin{document}


\title{On the nature of the prototype LBV AG~Carinae \\ II. Witnessing a massive star evolving close to the Eddington and bistability limits}


\author{J. H. Groh\altaffilmark{1}}
\author{D. J. Hillier\altaffilmark{2}}
\author{A. Damineli\altaffilmark{3}}

\altaffiltext{1}{Max-Planck-Institut f\"ur Radioastronomie, Auf dem H\"ugel 69, D-53121 Bonn, Germany; \email{jgroh@mpifr.de}}
\altaffiltext{2}{Department of Physics and Astronomy, University of Pittsburgh,
3941 O'Hara Street, Pittsburgh, PA, 15260, USA}
\altaffiltext{3}{Instituto de Astronomia, Geof\'{\i}sica e Ci\^encias 
Atmosf\'ericas, Universidade de S\~ao Paulo, Rua do Mat\~ao 1226, Cidade 
Universit\'aria, 05508-900, S\~ao Paulo, SP, Brasil}

\begin{abstract}

We show that the significantly different effective temperatures ($\teff$) achieved by the luminous blue variable AG Carinae during the consecutive visual minima of 1985--1990 ($\teff \simeq 22,800$~K) and 2000--2001 ($\teff \simeq 17,000$~K)  place the star on different sides of the bistability limit, which occurs in line-driven stellar winds around $\teff \sim 21,000$~K. Decisive evidence is provided by huge changes in the optical depth of the Lyman continuum in the inner wind as $\teff$ changes during the S Dor cycle. These changes cause different Fe ionization structures in the inner wind. The bistability mechanism is also related to the different wind parameters during visual minima: the wind terminal velocity was 2--3 times higher and the mass-loss rate roughly two times smaller in 1985--1990 than in 2000--2003. We obtain a projected rotational velocity of $220 \pm 50 \kms$ during 1985--1990 which, combined with the high luminosity ($\lstar=1.5\times10^{6} \lsun$), puts AG Car extremely close to the Eddington limit modified by rotation ($\Omega\Gamma$ limit): for an inclination angle of $90^\circ$, $\Gamma_\Omega \gtrsim1.0 $ for $M \lesssim 60~\msun$. Based on evolutionary models and mass budget, we obtain an initial mass of $\sim 100~\msun$ and a current mass of $\sim60-70~\msun$ for AG Car. Therefore, AG Car is close to, if not at, the $\Omega\Gamma$ limit during visual minimum. Assuming $M=70~\msun$, we find that $\Gamma_\Omega$ decreases from 0.93 to 0.72 as AG Car expands toward visual maximum, suggesting that the star is not above the Eddington limit during maximum phases.
\end{abstract}


\section{Introduction}  \label{intro}

Albeit rare, massive stars are the main contributors to the input of ionizing photons, energy, and momentum into the interstellar medium, and are responsible for a significant fraction of the chemical enrichment of their host galaxy. Massive stars evolve on relatively short timescales (a few $10^6$~years), and their evolution is strongly influenced by their strong mass loss and rotation \citep[e.g,][]{meynet05}.

Tremendous advancement in the understanding of the evolution of massive stars has been achieved in recent decades \citep{conti76,schaller92,meynet94,maeder94,langer94}. In particular, we now have insights into the effects of rotation on the
evolution of massive stars  \citep{maeder_araa00}. The models predict the existence of a short-lived, transitional stage, usually referred to as the Luminous Blue Variable (LBV) phase \citep{conti84,hd94}, during which the star has a high mass-loss rate ($\mdot \sim 10^{-5}-10^{-3}$ $\msunyr$). In the current picture of stellar evolution, LBVs are rapidly evolving massive stars in the transitory phase from being an O-type star burning hydrogen in its core to a hydrogen-poor Wolf-Rayet, helium-core burning star \citep{hd94,maeder_araa00,meynet00,meynet03}. The LBV phase is likely an unavoidable evolutionary stage which massive stars with initial mass $M_\mathrm{ini} \gtrsim 40~\msun$ will experience during their lifetime before becoming WRs and exploding as a supernova \citep{hd94}. Whether LBV progenitors have an upper limit for $M_\mathrm{ini}$, as suggested by the latest evolutionary models \citep{meynet03}, still remains to be seen. Several LBVs, such as Eta Car \citep{dh97,hillier01}, AG Car \citep{ghd09}, and the Pistol star \citep{figer98}, among others (see \citealt{humphreys78,hd94,vg01,clark05}), are known to have bolometric luminosities in excess of $10^6~\lsun$, which suggests a progenitor with $M_\mathrm{ini}\gtrsim100~\msun$.

Following recent determinations that the mass-loss rates of O-type stars are a factor of 3--20 lower than those adopted by the evolutionary models cited above \citep{crowther02,hillier03,massa03,kramer03,evans04,bouret03,bouret05,puls06}, and the detection of massive nebulae around LBVs, there is growing evidence supporting the idea that the evolution of massive stars is even more dominated by strong mass loss during the LBV phase than previously thought \citep{so06}. Surprisingly, some recent work suggests that some core-collapse SNe have LBV progenitors \citep{kv06,smith07,galyam07,trundle08}, which dramatically enhances the cosmological importance of LBVs and poses a great challenge to the current paradigm of massive star evolution. Some of these type-II SNe occurred just a few years after the progenitor suffered a giant mass ejection \`a la Eta Carinae in the 1840's \citep{smith07}, raising the following provocative hypothesis as to the near-term fate of Eta Car itself and other LBVs ---  LBVs might fail to lose additional significant amounts of mass and, thus, never become a Wolf-Rayet star \citep{smith07}. 

Only a dozen or so LBVs are known in the Galaxy \citep{hd94,vg01,clark05}, and determining their fundamental parameters and evolutionary status is a key to understanding unstable massive stars and Population III objects in the high-redshift Universe. 

AG Carinae (=HD 94910, $\alpha_{2000}$=10h56m11.6s, $\delta_{2000}$=-60d27min12.8s) is one of the brightest and most famous LBV stars. \citet{cannon16} classified AG Car as a ``P Cygni-type" star almost a century ago, while its
photometric variability was discovered by \citet{greenstein38}. Its high luminosity was first reported by \citet{thackeray50}, who also noticed the presence of a bipolar nebula around the central star.

After these pioneering works, intense observing campaigns in the past decades have provided a wealth of observational data on AG Car, showing that the star has photometric, spectroscopic, and polarimetric variability on timescales from days to decades (\citealt{leitherer94,hd94,stahl01,vg01,davies05}; see references in \citealt{ghd09}). The strongest variability occurs cyclically on timescales of decades and is known as the S-Dor cycle, during which the visual magnitude of the star decreases to $\sim6.0$ mag \citep{vg01}. 

The advent of non-LTE, full line-blanketed radiative transfer codes, such as {\sc CMFGEN} \citep{hm98}, has allowed us to obtain precise fundamental parameters for AG Car during the visual minimum epochs of the S-Dor cycle and to gain insights into the nature of the star (\citealt{ghd09}, hereafter Paper I).
\defcitealias{ghd09}{Paper I}

Following the determination of strikingly different stellar and wind parameters at different minima (\citetalias{ghd09}), 
we would like to address the following questions in the present paper. Is the bistability mechanism present in AG Car? What was the initial mass, and what is the  current mass of AG Car? Is AG Car close to the Eddington limit? If so, is it during visual minimum or on the way to visual maximum? How does rotation modify such a scenario? What is the present evolutionary status of AG Car?

This paper is organized as follows. In Sect. \ref{obsmod} we briefly describe {\sc CMFGEN}, the radiative transfer code employed in the analysis of the observed spectra of AG Car. A detailed comparison between the minimum phases of the S-Dor cycle of AG Car under the light of the bistability mechanism are considered in Sect. \ref{compbi}. In Sect. \ref{rot} we present evidence that AG Car was rotating fast during the visual minimum of 1985--1990. The evolutionary status, initial mass, and current mass of AG Car are considered in Sect. \ref{evol}. An analysis of how the fundamental parameters of AG Car change from visual minimum toward maximum of the S-Dor cycle is provided in Sect. \ref{modus}. The conclusions of this paper are summarized in Sect. \ref{conclusions}.

\section{Quantitative spectroscopic modeling }\label{obsmod}

The photometric and spectroscopic observations of AG Car during visual minimum were extensively described in \citetalias{ghd09} and references therein, to which we refer the reader for further details.

The detailed spectroscopic analysis of AG Car during visual minimum was done using the radiative transfer code {\sc CMFGEN} \citep{hillier87a,hillier90,hm98,hm99} and was presented in \citetalias{ghd09}. {\sc CMFGEN} has been successfully used to study other LBVs and related objects \citep{sc94,najarro94,najarro97,najarro01,hillier98,hillier01,hillier06,figer98,drissen01,bresolin02,marcolino07}. Since the code has been extensively discussed in the aforementioned references and in \citetalias{ghd09}, we concisely describe its main characteristics below.

CMFGEN assumes a spherical-symmetric, steady-state outflow and computes continuum and line formation in the non-LTE regime. Each model is defined by the hydrostatic stellar radius R$_\star$, bolometric luminosity L$_\star$, mass-loss rate $\dot{M}$, wind terminal velocity $\vinf$, stellar mass $M$, and abundances Z$_i$ of the included species. The radius of the hydrostatic core is defined as $v(\rstar)=v_\mathrm{sonic}/3$, in order to avoid any effects due to the strong wind in the determination of $\rstar$. The distance to AG Car cannot be constrained from the spectroscopic analysis using CMFGEN. Throughout this work, we assume that AG Car is located at $d=6~\mathrm{kpc}$ \citep{humphreys89,hoekzema92}. We refer to \citetalias{ghd09} for further discussion on the effects of the distance on the derived parameters.

CMFGEN does not solve the hydrodynamic equations to determine the wind structure, and therefore, a velocity law
$v(r)$ needs to be assumed a priori. In CMFGEN, the velocity law is parameterized by a beta-type velocity law which is modified at depth to smoothly merge with a quasi-hydrostatic structure at the sonic point. The velocity structure below the sonic point is iterated to fulfill the wind momentum equation with a maximum error of 10\% for $v \lesssim 8~\kms$. Systematic errors in the density structure below the sonic point might be present due to the proximity of the star to the Eddington limit and the choice of $M$, rotation rate, and viewing angle. 

CMFGEN allows for the presence of clumping within the wind. This is accomplished by assuming a homogeneous wind at distances close to
the star, and that clumps start to be formed at a given velocity $v_c$. The wind achieves a volume-filling factor $f$ at large distances, as follows: \begin{equation}
f(r)=f+(1-f)\exp[-v(r)/v_c]\,\,.
\end{equation} 

Full line blanketing is included consistently in CMFGEN through the concept of superlevels, which groups similar energy levels
into one single superlevel to be accounted for in the statistical equilibrium equations.  The final atomic model included H, He, C, N, O, Na, Mg, Al, Si, Ti, Cr, Mn, Fe, Co, and Ni.
Synthetic spectra were computed in the observer's frame using {\sc CMF\_FLUX} \citep{hm98,bh05}, including the effects due to rotation.

\section{The role of the bistability mechanism in AG~Car} \label{compbi}

After comparing the lightcurve and the spectroscopic evolution during both visual minimum phases \citepalias{ghd09}, it can be noted that the star presents
different characteristics in these epochs. An inspection of the AG Car lightcurve \citepalias{ghd09} shows that the visual minimum
phases of 1985--1990 and 2000--2003 are significantly different, especially regarding their duration. While the star was in a visual minimum phase for about 5 years between 1985--1990, the duration of the following visual minimum of 2000-2003 was much shorter, between 2 and 3 years \citepalias{ghd09}.

\begin{figure}
\resizebox{\hsize}{!}{\includegraphics{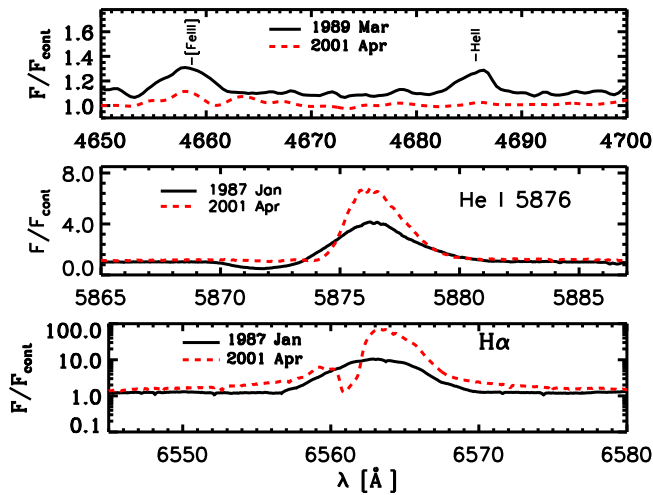}}
\caption{\label{comphotphases} Comparison between normalized observed line profiles of AG Car obtained during 1985--1990 and 2001 April, around the spectral lines of \ion{He}{2} 4686 (top panel), \ion{He}{1} 5876 (middle), and H$\alpha$ (bottom). The 2001 April observation in the upper panel was convolved with a Gaussian profile in order to match the resolution of the 1989 March data. The offset in the continuum level from 1.0 is due to the electron-scattering wings of N II lines at 4630--4653, \ion{He}{2} 4686, and \ion{He}{1} 4713. The data has $R\simeq 50,000$ in the middle and bottom panels and $R\simeq 3000$ in the upper panel.}
\end{figure}

There are also significant qualitative differences in the spectrum obtained at different visual minimum phases of AG Car, which can be inferred by comparing spectral lines of \ion{H}{1}, \ion{He}{1}, and \ion{He}{2}. Figure \ref{comphotphases} displays, for each visual minimum, the spectrum obtained during the epoch when the maximum value of $\teff$ was achieved, corresponding to 1985--1990 and 2000--2001. Figure \ref{comphotphases} presents the spectra obtained in 1987 June, 1989 March, and 2001 April, centered in three key lines for which high-resolution spectra were available: H$\alpha$, \ion{He}{1} 5876, and \ion{He}{2} 4686. H$\alpha$ emission, in both the line core and the electron-scattering wings, was much stronger in 2001 than in 1985--1990, indicating that the wind was denser in 2001. The P-Cygni absorption component of  H$\alpha$ was much deeper in 2001, while it was almost absent in 1985--1990. This indicates that the $n=2$ energy level population of H was higher in 2001, which was probably caused by the lower $\teff$ and, thus, increased amount of neutral H. We will come back to this point in Sect. \ref{ionouter}.
The \ion{He}{2} 4686 emission, on the other hand, was stronger in 1985--1990 than in 2001, confirming that $\teff$ was higher in 1985--1990 than in 2001. The line emission of \ion{He}{1} 5876 was stronger in 2001 than in 1985--1990, but this was due to the increased wind density and not because of a higher $\teff$, since the latter would also imply increased \ion{He}{2} 4686 emission, and the opposite is seen. The P-Cygni absorption profiles of H$\alpha$ and \ion{He}{1} 5876 both indicate that the wind terminal velocity was lower in 2001 than in 1985--1990.

The detailed spectroscopic analysis using CMFGEN models \citepalias{ghd09} confirms the above scenario and allows us to make quantitative conclusions through the comparison of the physical parameters obtained in the epochs when the maximum stellar temperature was achieved in consecutive visual
minimum phases. While $\lstar$ was similar and $\rstar$ was only 10\% different, three other physical parameters were significantly different;
namely, \teff, \mdot, and \vinf \citepalias{ghd09}. Notably, $\teff$ was approximately 5800 K higher in 1985--1990 than in 2001, while $\vinf$ was 2--3 times higher and $\mdot$ roughly two times smaller in 1985--1990 than in 2000--2003.

Based on our modeling, we suggest that the differences in the lightcurve and in the spectrum of AG Car during consecutive visual minimum phases are caused by the strikingly different {\it wind} parameters due to a change in the underlying stellar parameters \citepalias{ghd09}. {\em This is contrary to the current ideas about the nature of the S Dor cycles, where the visual minimum phase is representative of a quiescent phase of the LBV star. In this scenario, visual minimum phases should be equal to each other, which is not the case for AG Car.} Thus, the questions to be answered are: Why are the physical parameters different in consecutive visual minimum phases? Is there any intrinsic reason for this to occur?

The differences in $\teff$, $\mdot$, and $\vinf$ during the minimum phases of AG Car are immediately reminiscent of the bistability mechanism seen in line-driven winds \citep{pauldrach90,lamers95,vink99}, which has already been suggested to be responsible for the variable mass-loss rate seen in AG Car \citep{vink02}. Before going further and analyzing whether our results agree with previous works, we shall briefly summarize the physics behind the bistability mechanism.

\subsection{Background on the bistability mechanism}

There are two potential effects related to sudden changes in the stellar parameters, which are commonly referred to as bistability. 

First, according to the original proposition of \citet{pauldrach90} when analyzing models for the LBV P~Cygni, the classical bistability mechanism is characterized by a change in the optical depth of the Lyman continuum ($\taulyc$), which in turn changes the ionization stages of metals in the wind, causing a change in the amount of line driving, and thus, ultimately changing $\mdot$ and \vinf. Later, this concept was extended to a bistability {\it jump} by \citet{lamers95} to explain the physical parameters of blue supergiants, suggesting that the winds of those stars have two different regimes of $\mdot$ and \vinf, which switch abruptly from one to the other around 21,000~K. \citet{lamers95} proposed that blue supergiants with $\teff \leq 21,000~\mathrm{K}$ could typically have $\mdot$ 2 times higher (correspondingly, $\vinf$ is 2 times lower) than blue supergiants with $\teff \geq 21,000~\mathrm{K}$. The physical explanation for this phenomenon is a change in the ionization structure of iron in the inner stellar wind \citep{vink99}, which changes the number of spectral lines able to absorb/scatter radiation and, thus, able to drive the wind. This in turn affects the value of $\mdot$ and \vinf \citep{vink99}. 

While the \citet{pauldrach90} work referred to the wind of a single LBV star, \citet{lamers95} and \citet{vink99} were analyzing {\it different} OB supergiants. Both ideas were combined by \citet{vink02}, who suggested that the variable mass loss commonly inferred for LBVs is due to subtle changes in the line driving due to recombination/ionization of Fe$^{3+}$ into Fe$^{2+}$ and Fe$^{2+}$ into Fe$^{+}$. However, based on a detailed spectroscopic analysis with CMFGEN of a large sample of OB stars, the idea of a ``jump" in $\mdot$ and $\vinf$ of OB supergiants around $\teff \geq 21,000~\mathrm{K}$ has been disputed by \citet{crowther06}, who suggest a more continuous variation of  $\mdot$ and $\vinf$ with temperature.

Second, there is also the wind instability, which causes sudden changes in the H$\alpha$ line profiles, intensity of \ion{Fe}{2} lines, and $\taulyc$, for a relative small change in $\lstar$, $M$, and/or $\teff$. This effect is well known in LBVs and has been discussed in the literature for P Cygni \citep{najarro97}, HDE~316285 \citep{hillier98}, and Eta Carinae \citep{hillier01}. Interestingly, the wind instability and the classical bistability appear to occur at similar $\teff$, causing feedback effects as discussed by \citet{pauldrach90}.

\subsection{The behavior of the Lyman continuum of AG Car during visual minimum epochs \label{lycbeha}}
 
In order to check whether the bistability mechanism can explain the presence of different stellar and wind parameters in consecutive visual  minimum phases of AG Car, we first turn our attention to the underlying cause of the bistability mechanism, which is the change in $\taulyc$. We analyzed the behavior of $\taulyc$ in selected models obtained for both minima, which are shown in Fig. \ref{lymancontion}. While we obtained $\taulyc < 1 $  throughout the wind during 1985--1990 (i.e. optically thin), an increase by three orders of magnitude in $\taulyc$ was derived in models during 2000--2003, making the Lyman continuum completely optically thick. Therefore, we first conclude that {\it the different stellar parameters at the base of the wind of AG Car in different visual  minimum phases cause a huge change in $\taulyc$, which in turn cause significant differences in $\mdot$ and $\vinf$, implying a very different spectral morphology.} In particular, we conclude that when AG Car had $\teff \geq 19,500~\mathrm{K}$ (i.\,e.,  during 1985--1990), the Lyman continuum was optically thin, $\mdot$ was $\sim2$ times lower than in 2001, and $\vinf\sim3$  times higher than in 2001, when the Lyman continuum was optically thick and $\teff \simeq 17,000~\mathrm{K}$.

Our results confirm the findings from \citet{vink02} that AG Car lies on different sides of the bistability regime at different epochs \citep{pauldrach90}. We found that between 1985--1990, AG Car had $\teff \simeq22,800$ K, which we interpret as being on the hot side of the bistability regime, while in 2001 AG Car had $\teff\simeq17,000$ K, which we interpret as the star crossing the bistability limit into the cool
side. This result implies that the effective temperature for occurrence of the bistability in AG Car is somewhere between 17,000 K and 22,800 K. Based on our detailed CMFGEN spectroscopic analysis, we suggest that this limit is closer to the earlier, around 17,000K -- 18,000 K, because in this case, small changes of 5\% in \lstar, \mdot, and/or $\rstar$ caused significant changes in $\taulyc$, in the ionization structure of the wind, and in the emerging spectrum. 

\subsection{The iron ionization structure in the inner wind of AG Car \label{ironion}}

We shall now discuss the iron ionization structure predicted by our CMFGEN modeling in the inner wind of AG Car, since the huge change in $\taulyc$ is expected to impact the iron ionization structure \citep{vink02}. 

The inclusion of several physical processes in the current generation of radiative transfer stellar wind codes such as CMFGEN turned out to be crucial for the analysis of the ionization structure of the wind of early-type stars. In particular, the exact treatment of radiative transfer in the co-moving frame, the inclusion of full line blanketing, wind clumping, and consistent modeling of the wind and photosphere have all had a huge effect on the ionization structure of early-type stars (e.g \citealt{hm99,crowther02,hillier03,bouret05,martins05,puls06,hamann06,crowther07}). We anticipate that the ionization structure of LBV stars will also be affected. 

\begin{figure}
\resizebox{\hsize}{!}{\includegraphics{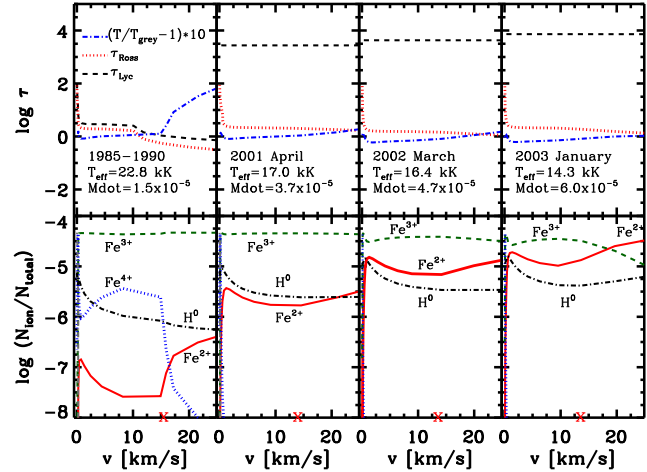}}
\caption{\label{lymancontion}  {\it Top}: Optical depth of the Lyman continuum (black dashed line), Rosseland optical depth (red dotted), and ratio between the wind temperature and the grey temperature (blue dot dashed) as a function of velocity, derived from CMFGEN models of AG Car for the 1985--1990 and 2000--2003 minima. The models are ordered by decreasing $\teff$ (and increasing $\mdot$) from left to right. {\it Bottom}: Corresponding ionization structure of Fe and H of the inner wind of AG Car as a function of velocity. Since H is mainly ionized in the inner wind of AG Car at all epochs shown in this Figure, only the H$^0$ abundance is shown for clarity. For each model, the velocity of the sonic point is indicated by a cross at the horizontal axis.}
\end{figure}

The Fe ionization structure of AG Car during 1985--1990 and 2000--2003 obtained from the CMFGEN models is presented in the bottom row of Figure \ref{lymancontion}. We noticed that the ionization structure of Fe before the sonic point, which is around $15 \kms$ for all the above models and below which $\mdot$ is set, is modified when $\taulyc$ changes. During 1985--1990 the dominant ionization stage was Fe$^{3+}$, while in the following visual minimum phase of 2001 there is a steady increase in Fe$^{2+}$ as $\teff$ decreased. Since Fe$^{2+}$ has a higher opacity than Fe$^{3+}$, particularly in the region where the stellar flux is high, more momentum can be removed from the radiation field and transferred to the wind, causing a higher mass-loss rate. Nevertheless, as seen in Fig. \ref{lymancontion}, Fe$^{3+}$ is still the dominant ionization stage below the sonic point in the temperature regime found for AG Car during visual minimum.

\subsection{Is there a {\it jump} in the wind properties of LBVs due to the bistability mechanism?}

\citet{vink02} proposed that the bistability mechanism is the cause of the variable behavior of the mass-loss of LBVs as a function of the effective temperature, predicting the presence of jumps in the $\mdot$ vs. $\teff$ relationship at $\teff \sim 21,000~\mathrm{K}$ and $\teff \sim 10,000~\mathrm{K}$ due to the recombination of Fe$^{3+}$ into Fe$^{2+}$ and Fe$^{2+}$ into Fe$^{+}$, respectively.

\begin{figure}
\resizebox{\hsize}{!}{\includegraphics{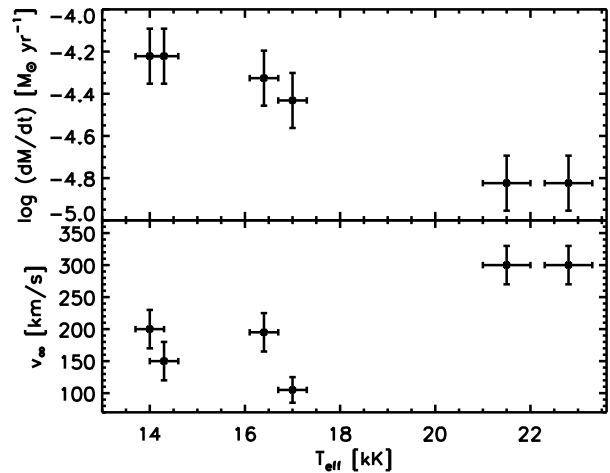}}
\caption{\label{mdotvinfmin} Mass-loss rate (upper row) and wind terminal velocity (bottom row) of AG Car as a function of $\teff$ during visual minimum phases. }
\end{figure}

Figure \ref{mdotvinfmin} presents $\mdot$ as a function of $\teff$ during the visual minimum epochs of AG Car analyzed in Paper I. Is there clear evidence of a jump in $\mdot$ as a function of $\teff$ as AG Car crosses from one side of the bistability regime to the other? From the point of view of the spectroscopic modeling, we cannot give a definitive answer to this question for the following reasons. In general, we can see a progressive increase in $\mdot$ when $\teff$ decreases in the range $21,000-14,000~\mathrm{K}$. Close to the region where the Lyman continuum becomes optically thick (aka the bistability limit) at $\teff \simeq 17,000-18,000~\mathrm{K}$, we can identify AG Car in the two regimes predicted by \citet{vink02}. On the hot side, $\mdot$ is lower than on the cool side, while $\vinf$ is higher. However, whether the star switches abruptly from one regime to another or changes gradually cannot be determined since AG Car has never been observed at temperatures very close to the bistability limit. Long-term spectroscopic monitoring has shown that $\teff$ of AG Car decreases quickly  from visual minimum to maximum (on timescales of a few months, \citealt{leitherer94,stahl01}; Paper III), and monitoring AG Car in intervals of weeks would be required. Even then, our ability to derive the stellar and wind parameters would likely be hampered by strong time-dependent effects, since the flow timescale during these epochs is also of the order of a few months. Nevertheless, we can conclude from the short timescale that the stellar wind is highly unstable when close to the bistability limit, and the inner layers might be exposed to some sort of feedback from the unstable wind \citep{pauldrach90}. 

The differences in our results from the predictions of \citet{vink02} are likely a combination three factors: a) the inclusion of full line blanketing and wind clumping in our analysis, b) the higher $L$ derived in our work compared to the value assumed by \citet{vink02}, and c) due to the fact that the ratio $\vinf/\vesc$ is definitely not constant when $\teff$ changes, contrary to the assumption of \citet{vink02}. Furthermore, $\mdot$ and $\vinf/\vesc$ of AG Car are greatly modified by rapid rotation \citep{ghd06}, a phenomenon which was not included in the \citet{vink02} models. The influence of rotation will also be variable as a function of $\teff$, since the ratio $\vrot/\vcrit$ changes from visual minimum toward maximum (\citealt{ghd06}; see Sect. \ref{rot}).   

The behavior of $\vinf$ as a function of $\teff$ is more irregular \citepalias[see ][]{ghd09}. First, there is a decrease in $\vinf$ from $\teff \simeq 23,000~\K$ until $\teff \simeq 17,000~\K$. Then, a steep increase until $\teff \simeq 16,400~\K$ is seen, and a shallow decrease again until  $\teff \simeq 14,000~\K$. While the overall behavior of $\vinf$ might be explained in terms of the bistability mechanism through changes in the line driving \citep{vink02}, there are a number of physical mechanisms which likely have a large impact on the behavior of $\vinf$  on the cool side: the fast rotation derived in AG Car, the presence of a latitude-dependent wind and departure from spherical symmetry, the accuracy of the velocity structure near the sonic point, and time-dependent effects. 

\subsection{The H and Fe ionization structures in the outer wind of AG Car and the variable P-Cygni absorption profile \label{ionouter}}

\begin{figure}
\resizebox{\hsize}{!}{\includegraphics{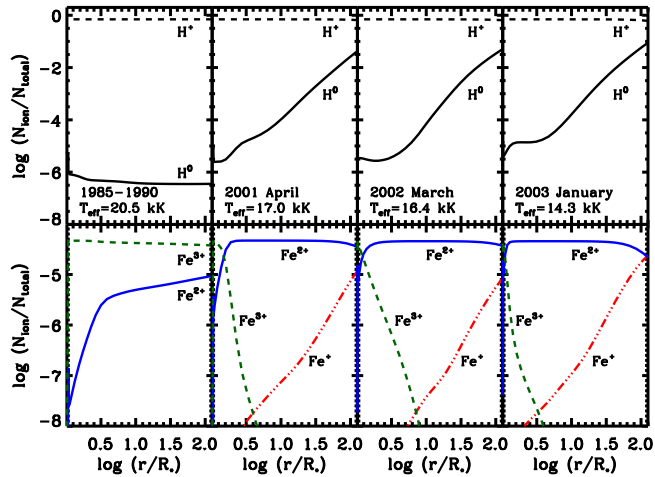}}
\caption{\label{agcionesttot}Ionization structure of H (upper row) and Fe (bottom row) as a function of distance in the wind of AG Car during visual minimum ($\teff$ decreasing from left to the right). The region shown in Fig. \ref{lymancontion} corresponds roughly to $ 0 < \log(r/\rstar) < 0.1$. }
\end{figure}

The presence of strong \ion{Fe}{2} lines in the spectrum of AG Car during the visual minimum of 2000--2003 (which were absent during 1985--1990) provide solid evidence that the ionization structure of the {\it outer} wind of AG Car was also different in those minima.  Figure \ref{agcionesttot} presents the  H and Fe ionization structures in the whole wind of AG Car during visual minimum phases as derived from the CMFGEN models with the parameters from \citetalias{ghd09}. It can be seen that in addition to the recombination of Fe$^{3+}$ into Fe$^{2+}$ in the inner wind during the 2000--2003 visual minimum, there is also recombination of  
H$^{+}$ into H$^{0}$ and of Fe$^{2+}$ into Fe$^{+}$ in the outer wind of AG Car. This explains the appearance of P-Cygni absorptions during the latter visual  minimum, which is only possible when a significant ($10^{-3}$) fraction of hydrogen is recombined. The same mechanism explains the
absence of P-Cygni absorption during 1985--1990, since hydrogen was ionized throughout the whole wind. In addition, as first noted by
\citet{hillier98} and \citet{hillier01}, the Fe ionization structure in the outer wind is dominated by the following charge-exchange reaction with
H, \begin{equation}
\mathrm{Fe^{2+} + H \leftrightarrow Fe^{+} + H^{+}\,\,.}
\end{equation} This reaction implies that hydrogen recombination is also responsible for the recombination of Fe$^{2+}$ into Fe$^{+}$, thus explaining why there are no strong Fe II lines in the 1985--1990 spectrum while they are relatively strong in data from 2000--2003.

The marked difference in the ionization structure of the outer wind during different visual minimum phases of AG Car led us to investigate whether the recombination of hydrogen and Fe$^{2+}$ into Fe$^{+}$ in the outer wind and the consequent appearance of a P-Cyg absorption component are also due to the bistability mechanism (i.e. change in the optical depth of the Lyman continuum).

We found that when $\taulyc$ increases, the relative population between the hydrogen levels 2 and 1 also increases, and so does the optical depth of the Balmer lines, such as H$\alpha$ and H$\beta$, explaining the increase in the absorption of those lines. 

Interestingly, variable P-Cygni absorption has been observed in the LBV candidate He 3-519 \citep{sc94,crowther97}, which might indicate that He 3-519 is also close to the bistability limit. Further work is highly desirable in order to prove whether this speculation holds for LBVs in general.

\section{Rotation}\label{rot}

The analysis of broad absorptions due to \ion{Si}{4} 4088--4116 {\AA} in the high-resolution AG Car spectra obtained during the 2000--2003 visual minimum has shown direct evidence, for the first time, that some LBVs rotate at a high fraction of their critical rotational velocity \citep{ghd06}. Using a similar methodology, the bona-fide LBV HR Carinae was also shown to possess a rotational velocity close to critical \citep{gdh09}. Since the \ion{Si}{4} lines were also present in AG Car during the 1985--1990 visual minimum \citep{stahl86,sc94,wf2000}, we analyzed whether a similar fast rotation could be noticed during these epochs. Exactly the same technique as described in \citet{ghd06} was employed. 

\begin{figure}
\resizebox{\hsize}{!}{\includegraphics{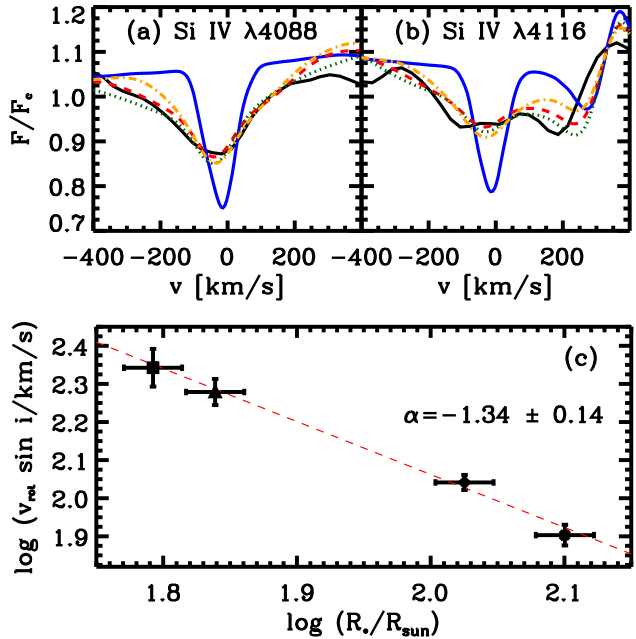}}
\caption{\label{agcvrotr}{\it a)}  Medium-resolution
\ion{Si}{4} 4088 absorption profile of AG Car observed in 1989 March (solid black line) compared to a set of CMFGEN models computed without
rotation (solid grey line) and with a projected rotational velocity  of $150~\kms$ (dot-dashed 
line), $220~\kms$ (dotted line), and $270~\kms$ (dashed line). Despite the  medium resolution of $\sim 100~\kms$ of the observations, 
we can clearly infer the presence of a high rotational velocity in AG Car in this epoch. {\it b)} Same as panel {\it a}, but for \ion{Si}{4} 4116. {\it c)} Rotational velocity of the surface of AG Car as a function of $\rstar$ in log--log scale.}
\end{figure}

Although the spectra of AG Car obtained in 1989 March have only moderate resolution around the \ion{Si}{4} lines (R$\sim$100 \kms), significant evidence of fast rotation  can be inferred from the comparison between the observed spectrum and the 2-D models presented in Fig. \ref{agcvrotr}. The \ion{Si}{4} line profiles from models without rotation, convolved to the same spectral resolution as the observations, are narrower and have stronger central absorption than the observed \ion{Si}{4} line profiles of AG Car from 1989 March. {\it  This is clear evidence that fast rotation is present.} The best fit was achieved using $\vrot\, {\mathrm \sin i}=220 \pm 50~\kms$, mainly using \ion{Si}{4} 4088 as a diagnostic (Fig. \ref{agcvrotr}a,) since \ion{Si}{4} 4116 is severely blended with \ion{He}{1} 4120 at the data resolution (Fig. \ref{agcvrotr}b). We estimate that $\vrot$ likely had similar values during the whole 1985--1990  visual minimum, since there was little variability in the stellar parameters during these epochs \citepalias{ghd09}. Indeed, a tracing of a medium resolution photographic-plate spectrum of AG Car obtained in 1985 \citep{stahl86} shows broad \ion{Si}{4} 4088--4116 absorptions which are very similar to those predicted by our rotating models.

Figure \ref{agcvrotr}c presents the rotational velocity of AG Car as a function of $\rstar$ for each epoch. Using a least-squares linear fit, we obtained $v_{\mathrm{rot}} \propto \rstar^{-1.34 \pm 0.14}$, where the quoted error only includes the statistical errors. Additional systematic errors, which are currently hard to estimate, might be present due to model assumptions. A flatter dependence of $\vrot$ on $\rstar$ might be obtained when including the deformation of the stellar surface from spherical symmetry, variation of the stellar and wind parameters as a function of latitude, and a more complex treatment of the density structure at and below the sonic point in the model. As a consequence, we cannot exclude that $v_{\mathrm{rot}} \propto \rstar^{-1}$. In addition, our result strongly suggests that magnetic fields are not important on the surface of AG Car; otherwise, significant deviations from the above relationship would be expected  (A. Maeder, private communication). If the magnetic field on the hydrostatic surface of AG Car played an important role, one would expect that the magnetic field would maintain the surface material corotating out to some distance. In this case, the surface rotational velocity would behave as that of a solid body, i.\,e., increasing with $\rstar$. This is contradictory with our findings.

The rotational period of AG Car during 1985--1990 is $\mathrm{P_{rot}}=13 \pm 2 $ days, assuming $\rstar=62~\rsun$ and $\vrot=220~\kms$. LBV stars have low-amplitude photometric variability (the so-called ``micro-variability") in timescales of days to weeks \citep{lamers98,vg01}. Generally, the currently accepted mechanism for such behavior is non-radial pulsation \citep{lamers98}. In the case of AG Car, the micro-variability period during 1985--1990 was between 11--14 days \citep{vg88}.

The similarity between the values of $\mathrm{P_{rot}}$ and the micro-variability period is striking. We suggest that they might be related to each other and that the micro-variability is modulated by the rotational period of AG Car, which would be the case if spots are present on the stellar photosphere, for instance.  Since the stellar photosphere of AG Car is extended and located in the stellar wind, the presence of spots would suggest that structures propagate from the stellar surface into the wind, possibly causing effects on line profiles. If the photometric micro-variability is indeed modulated by rotation, we anticipate that the micro-variability period should be proportional to $\rstar^{-2.34}$ in the case of AG Car, since $\vrot$ is proportional to $\rstar^{-1.34}$ for AG Car. 

If rotation is somehow related to the micro-variability of LBVs, in principle, their rotational velocities could be derived from the micro-variability period once
the stellar radius $\rstar$ is known. This analysis is beyond the scope of this paper and is the subject of an ongoing work.  It is
interesting to note that for other LBVs such as HR Car and S Dor, the variations in the micro-variability period between visual minimum and maximum
is about 5 \citep{lamers98}, which is consistent with a rotational modulation.

\section{Evolutionary status of AG Car} \label{evol}

\subsection{Initial mass and location of AG Car in the HR diagram}

In this Section we compare the fundamental parameters obtained for AG Car \citepalias{ghd09} with the evolutionary tracks from \citet{meynet03} (hereafter MM03). There are a couple of caveats which need to be considered: 1) the problematic definition of $\teff$ in LBVs and WRs because of the presence of an optically-thick wind and 2) the strong, long-term variability seen in AG Car during the S-Dor cycle.
\defcitealias{meynet03}{MM03}

The first issue is a direct consequence of the high mass-loss rate of LBVs and WRs, which causes the presence of an extended photosphere. The hydrostatic base of the wind ($\rstar$ in our definition) has a higher temperature ($\tstar$) than that of the photosphere ($\teff$, defined at $\tau_\mathrm{Ross}=2/3$). The relative decrease of $\teff$ compared to $\tstar$ depends on how optically thick the wind is, which is a function of $M$, $\mdot$, and wind velocity law (e.g., \citealt{langer89}).  Additional complications come from the presence of wind clumping and time-dependent effects in LBVs, which are both not considered by evolutionary models.

Since the \citetalias{meynet03} evolutionary models do not solve the radiative transfer in the extended stellar atmosphere as CMFGEN does, they cannot self-consistently compute $\teff$ during the LBV and WR phases. Only a rough estimate of the effect of the optically-thick wind on $\teff$ can be calculated based on the stellar parameters and on the predicted mass-loss rate \citep{langer89,schaller92,mm05}. Obviously, that result will only be approximate and might not be accurate, as any changes in the assumed parameters (such as $\mdot$) will bias the comparison between $\teff$ determined from the CMFGEN models and $\teff$ estimated by the evolutionary models. {\it Therefore, comparing the value of $\teff$ predicted by the evolutionary models with values obtained from the atmospheric radiative transfer codes might be misleading in the case of stars with dense winds such as LBVs and WRs.} In this case, stars with a similar $\teff$ can have very different underlying stellar parameters at the base of the wind if they have different $M$, $\mdot$, and/or $\vinf$.

The second issue is related to the strong variability of AG Car due to the S Dor-type variability. Which epoch of the S-Dor cycle should be compared with the evolutionary tracks, since the models do not account for the S-Dor type variability? The answer to this question is not obvious. Although the current ideas about the S-Dor cycle consider the visual minimum phase the quiescent phase, followed by eruptions during maximum (e.g. \citealt{hd94}), there is much evidence which might not corroborate this point. First, we detected that $\Omega\Gamma$ in AG Car is higher during visual minimum than during maximum (Sect. \ref{modus}). Therefore, if AG Car is to settle in quiescence during visual minimum, it has to do so extremely close to the Eddington limit. Second, even if we assume that the star is at quiescence during visual minimum, AG Car can have different stellar parameters in different minimum phases \citepalias{ghd09}. Which one should be compared with the evolutionary tracks? Third, some well-documented LBVs, such as S Doradus \citep{vg97b} and R127 \citep{walborn08}, have been observed at maximum for extended time intervals. Fourth, there is historical evidence that some LBVs, such as AG Car, have presented the S-Dor type instability for over a century \citep{vg01}, and the key point is how long the S-Dor cycles are present. {\it What if the S Dor-type instability is present during the majority of the LBV phase for these stars?} In this case we would just be witnessing these unstable, massive stars evolving in real time, and they would never really be in quiescence during the LBV phase. A definitive answer to this point can only be provided when the physical mechanism behind the change of the hydrostatic radius of LBVs, which ultimately causes the S Dor-type variability, is recognized and included in the stellar evolutionary models.

\begin{figure}
\resizebox{\hsize}{!}{\includegraphics{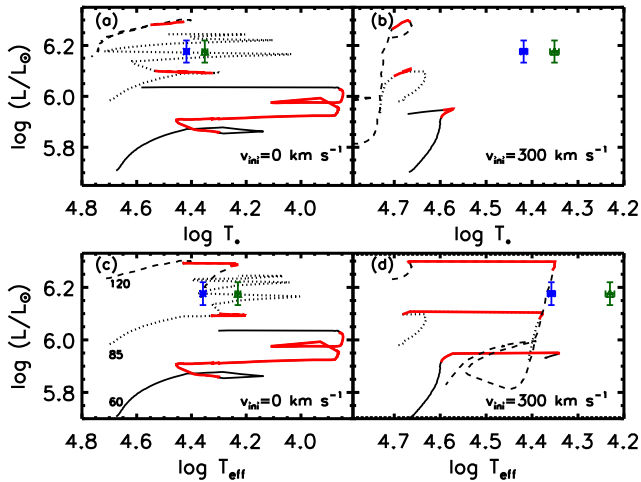}} 
\caption{\label{agcevoltrack} Position of AG Car in the HR diagram based on the physical parameters derived in \citetalias{ghd09} for the visual minimum of 1985--1990 (x symbol) and 2000--2003 (open triangle).  Evolutionary tracks from \citetalias{meynet03} for non-rotating stars ({\it a} and {\it c}) and stars with an initial rotational velocity of 300 \kms ({\it b} and {\it d}) are shown for initial masses of $60~\msun$ (full line), $85~\msun$ (dotted), and $120~\msun$ (dashed). The upper panels ({\it a} and {\it b}) present the value of $\tstar$ from the CMFGEN models with the effective temperature predicted by the \citetalias{meynet03} {\it not} corrected by optical-depth effects, while the lower panels ({\it c} and {\it d}) compare the value of $\teff$ from CMFGEN with the values from the  \citetalias{meynet03} evolutionary models corrected for optical-depth effects due to the presence of a dense wind. The thick part of each track corresponds to epochs where the helium abundance is $0.5 < Y < 0.7$, which is the range commonly found in LBVs.}
\end{figure}

Keeping these caveats in mind, we present in Figure \ref{agcevoltrack} the location of AG Car in the HR diagram during the visual  minimum phases of 1985--1990 and 2000--2001 based on the parameters derived in \citetalias{ghd09}, compared to those predicted by the evolutionary tracks for non-rotating and rotating stars from \citetalias{meynet03}. The He abundance of AG Car (Y=0.62, \citetalias{ghd09}) is a strong constraint on the evolutionary stage of AG Car, restricting the comparison to a very specific part of the evolutionary tracks. According to them, the bolometric luminosity of AG Car is consistent with an initial mass of $\sim100~\msun$, independent of the initial rotational velocity\footnote{Rotating models are slightly more luminous than non-rotating models \citepalias{meynet03}, but not sufficiently enough in order to change this conclusion.}.

Technically, only the \citetalias{meynet03} non-rotating models in the range $40\lesssim M \lesssim 60~\msun$ ($25\lesssim M \lesssim 45~\msun$ for rotating models) have a bona-fide LBV phase; in models with $M \gtrsim60~\msun$ the star skips the LBV phase and goes directly to the WR phase \citepalias{meynet03}. While both the rotating and non-rotating models can reproduce fairly well the effective temperature of AG Car (Fig. \ref{agcevoltrack}, lower panels), only the non-rotating model is consistent with the value of $\tstar$ derived for AG Car (upper panels). The evolutionary models with rotation predict a much hotter ($\tstar\simeq49,000~\K$) WN star at the base of the wind, corresponding to a WN7 spectral type \citep{crowther07}. Even though the stellar evolution model with rotation can fairly reproduce the $\teff$ of AG Car, it cannot reproduce the value of $\tstar$ that we inferred based on the spectroscopic analysis. Therefore, based on current evolutionary models, we can rule out that the progenitor of AG Car had a high initial rotational velocity. 

The non-rotating models simultaneously reproduce the values of $\tstar$, $\teff$, and He/H; the value of $\tstar$ is consistent with a WN11 spectral type. However, the fast rotation of AG Car is puzzling in the non-rotating scenario. One possibility is that the initial rotational velocity of AG Car was significantly lower than 300~\kms. Another possibility is that additional processes need to be taken into account by the evolutionary models in order to fully explain the stellar properties during the LBV phase.  For instance, it would be desirable to include giant outbursts of several $\msun$ in the evolutionary models to investigate the effects of this event on the evolution of an LBV. 

\subsection{Current mass}
\subsubsection{The minimum mass according to the Eddington limit} \label{massdep}

The influence of the stellar mass $M$ and effective gravity $g$ on the spectrum of OB supergiants is usually obtained from the broadening of H and He lines. However, the photosphere is extended when a dense stellar wind is present and, as the location of the photosphere approaches the sonic point, line emission from the wind fills out the original photospheric absorption profile. Thus, in the case of  LBVs and WR stars, the dense wind causes the photosphere to be located at velocities larger than the sonic point and strong emission lines appear in the spectrum. Therefore, the influence of log $g$ through the analysis of hydrogen absorption lines is not applicable to AG Car.

To analyze the influence of log $g$ on the spectrum of AG Car, we ran CMFGEN models assuming different stellar masses in the range $40~\msun < M < 150~\msun$, keeping the other model parameters constant,  such as a He mass fraction of 0.62 \citepalias{ghd09}. The stellar mass affects the Eddington parameter $\Gamma$ since
\begin{equation}
\label{gamma}
\Gamma=\kappa \lstar / (4\pi G c M)\,\,,
\end{equation}
where $\kappa$ is the total wind opacity and the other symbols have their usual meaning. For epochs where the wind density was large enough that $\vphot > \vs$ and the star was not so close to the Eddington limit ($\Gamma \lesssim 0.7$), such as during 2002 March -- 2002 July and 2003 January, we found little influence of log $g$ (and hence $M$) on the emerging spectrum. The physical parameters derived from  models with different log $g$ were virtually the same, and little insight on the value of $M$ could be obtained. This result is similar to what has been found for Eta Car \citep{hm98} and WR stars in general.

However, for epochs when $\Gamma \gtrsim 0.8 $ (1985--1990 and 2000--2001), we found that the emerging spectrum changed significantly as a function of the adopted $M$, and so did the parameters obtained from the spectroscopic analysis. Unfortunately, $M$ could not be directly derived from the models, since a reasonable fit of the spectrum could still be obtained by adjusting other parameters such as $\teff$, $\mdot$, and $f$. We derived that the changes in $M$ affected mostly $\teff$, which varied between 19600--23800 K when $M$ varied from 40 to 150 \msun, as shown in Fig. \ref{gammamass}. The value of $\teff$ is sensitive to the density and velocity structure below the sonic point, and we determined that $\teff$ is usually 1000--2000 K higher in models which adopted a fixed isothermal scale height to match a hydrostatic structure (using a procedure described by \citealt{hillier03}).  The dependence of $\teff$ on $M$ in this parameter regime can be understood as being due to the formation of a progressively more extended photosphere as the star approaches the Eddington limit. When $M$ decreased (while $\lstar$ and $\teff$ were kept constant) we found that higher ionization was achieved in the wind. Therefore, a decrease in $\teff$ is needed to fit the observed spectrum again. 

The middle panel of Fig. \ref{gammamass} shows $\Gamma$, obtained from our CMFGEN modeling, as a function of $M$, computed in two locations of the wind: at high optical depth  ($\tau=100$ and $v\sim 0.03 \kms$) and at $\rstar$ ($v \simeq 5 \kms$). According to our CMFGEN models, AG Car has to have $M > 40~\msun$ to ensure $\Gamma<1$. Different relationships were obtained: while $\Gamma \propto M^{-0.96 \pm 0.09}$ at $v\simeq 5 \kms$, a much less steep variation was obtained at the base of the wind, at $v \simeq 0.03 \kms$ ($\Gamma \propto M^{-0.37 \pm 0.02}$).  The different steepness at different depths in the wind reflect the different photospheric extensions and  opacities as a function of $M$.

\subsubsection{Effects of fast rotation on the minimum mass of AG~Car: the $\Omega\Gamma$ limit} \label{roteff}

\defcitealias{mm_omega00}{MM00}

The high rotational velocity derived for AG Car during visual minimum (\citealt{ghd06}; Sect. \ref{rot}) will reduce the maximum luminosity that can be reached before $\Gamma = 1$ (\citealt{mm_omega00}, hereafter MM00) and will require a larger value of $M$ in order for the star to be below the Eddington parameter modified by rotation ($\Gamma_\Omega$). Here we show that, making a few reasonable assumptions, we are able to obtain a lower limit for $\Gamma_\Omega$ and $M$ when taking rotation into account.  

According to \citetalias{mm_omega00}, 
\begin{equation}
\label{gammaomega}
\Gamma_\Omega=\frac{\kappa(\theta) \lstar [1+\zeta(\theta)]}{4 \pi c G M \left(1-\frac{\Omega}{2 \pi G \rho_\mathrm{m}}\right)}\,\,,
\end{equation}
where $\theta$ is the latitude angle and $\zeta$ is a correction factor to the von Zeipel gravity-darkening law  to take into account the baroclinicity of the star \citepalias{mm_omega00}.

Since the stellar parameters of AG Car were obtained in \citetalias{ghd09} through a one-dimensional radiative transfer analysis, and the high rotational velocity will cause deviations from spherical symmetry, there are a few caveats which need to be taken into account to obtain a reasonable value of  $\Gamma_\Omega$.

First,  the mass-absorption coefficient and radiative flux may vary as a function of latitude in a fast-rotating star. Unfortunately, given the 1-D nature of our radiative transfer analysis, we cannot infer directly how these changes will impact the estimation of  $\Gamma_\Omega$. To obtain a rough estimate of how close AG Car is to the Eddington limit modified by rotation (the so-called $\Omega\Gamma$-limit), we ignore latitudinal variations of the mass-absorption coefficient and radiative flux in Eq. \ref{gammaomega}, which then reduces to
\begin{equation}
\label{gammaomega_1d}
\Gamma_\Omega \simeq \frac{\Gamma}{1-\frac{\Omega}{2 \pi G \rho_\mathrm{m}}}\,\,.
\end{equation}
To properly include realistic latitudinal variations of the opacity and radiative flux, one would need detailed hydrodynamical modeling coupled with radiative transfer calculations in 2-D, which is certainly well beyond the scope of this paper.  

At first glance, the assumption that the derived $\lstar$ corresponds to the equatorial value of $\lstar$ might imply that we are neglecting gravity-darkening, i.e, that the flux of a rotating star is larger at the polar direction than at the equator \citep{vz24}. However, the value of $\lstar$ of a rotating star derived using a 1-D model also depends on the viewing angle \citep{collins63,collins65,gillich08}. For instance, for a Main-Sequence O-star with $\teff=35,000~\K$ rotating at 0.9 of its critical velocity for break-up, the actual value of $\lstar$ is about 25\% higher than the derived value of $\lstar$ when observing equator-on and using a 1-D model (Warren \& Hillier, private communication). The presence of the wind will likely change this dependence. Taking into account that AG Car is seen nearly equator-on \citep{ghd06}, $\lstar$ obtained in \citetalias{ghd09} is  likely lower limit, and it is reasonable to assume that the flux at low latitudes is not much different from the one predicted by a spherical symmetric model with $\lstar=1.5\times10^{6}\lsun$ as in  \citetalias{ghd09}.

Equation \ref{gammaomega_1d} can be rewritten as a function of  $\vrot$ as \citepalias{mm_omega00}:
\begin{equation}
\label{gammaomega_1d_vrot}
\Gamma_\Omega \simeq \frac{\Gamma}{1- \left[\frac{4}{9} \frac{v_\mathrm{rot}^2}{v_\mathrm{crit,1}^2}  V'   \frac{R_\mathrm{pb}^2}{R^2_\mathrm{e}} \right]}\,\,,
\end{equation}
where $R_\mathrm{e}$ is the equatorial radius at a given $\vrot$, $R_\mathrm{eb}$ and $R_\mathrm{pb}$ are respectively the equatorial and polar radius at the classical break-up, $v_\mathrm{crit,1}=(GM/R_\mathrm{eb})^{0.5}$ is the first critical velocity for rotation, and $V'$ is the volume of the star relative to that of a sphere with  $R_\mathrm{pb}$,  as in \citetalias{mm_omega00}. Assuming that the factor $V'   \frac{R_\mathrm{pb}^2}{R^2_\mathrm{e}}$ is close to 1 in AG Car\footnote{\citetalias{mm_omega00} show that this factor varies between 1 and 0.813, and our test calculations show this produces a negligible effect compared to the other errors associated with the analysis.}, and using the definition of $v_\mathrm{crit,1}$, Eq. \ref{gammaomega_1d_vrot} reads as
\begin{equation}
\label{gammaomega_1d_vrot_simpl}
\Gamma_\Omega \simeq \frac{\Gamma}{1- \left[\frac{2}{3} \frac{v_\mathrm{rot}^2 R_\mathrm{pb}}{G M} \right]}\,\,.
\end{equation}

\begin{figure}
\resizebox{\hsize}{!}{\includegraphics{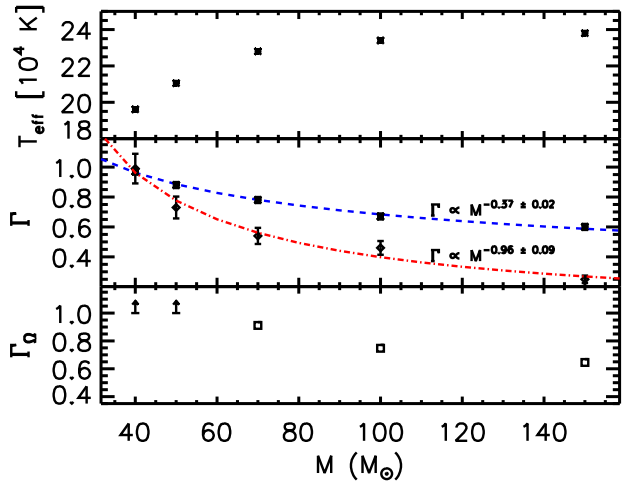}} 
\caption{\label{gammamass} {\it Top:} Derived value of $\teff$ as a function of  the stellar mass for different CMFGEN models for the 1985--1990 visual  minimum phase.
{\it Middle:} Eddington parameter $\Gamma$ as a function of mass for different CMFGEN models for the 1985--1990 visual minimum phase. Note that the $\teff$ and $\mdot$ of each model were also changed in order to fit the spectrum. The different symbols correspond to values of $\Gamma$ obtained at high optical depth ($\tau=100$ and $v\sim 0.03 \kms$ (asterisks) and at $v \simeq 5 \kms$  (open circles). A least-squares linear fit is shown for each sample. {\it Bottom:} Eddington parameter modified by rotation $\Gamma_\Omega$ as a function of the stellar M computed at the base of the wind at $v\sim 0.03 \kms$ (open squares). The arrows pointing upwards indicate that $\Omega \Gamma > 1 $, which was obtained for $M < 60 \msun$.}
\end{figure}

The second caveat regards how $\rstar$, determined from the 1-D radiative transfer model, relates to $R_\mathrm{pb}$. We argue that $\rstar \simeq R_\mathrm{pb}$ because 1) $\vrot / v_\mathrm{crit,1}$ is at maximum 0.7 for AG Car, meaning that the shape of the star does not significantly departs from spherical symmetry, and 2) the effects of fast rotation on the stellar structure were not included in CMFGEN in the computation of the hydrostatic atmospheric structure below the sonic point.  Using this simplification, we can then replace $R_\mathrm{pb}$ by $\rstar$ and, using $\Gamma$ and $\vrot$ determined from the spectroscopic analysis and assuming $i=90^\circ$, obtain $\Gamma_\Omega$ from models with different $M$ using Eq. \ref{gammaomega_1d_vrot_simpl}. The results from this calculations are shown in the bottom panel of Figure \ref{gammamass}.  

Since $\vrot$ is considerable,  the minimum mass of AG Car will be larger than the value of $\simeq 40~\msun$ derived in Section \ref{massdep} according to the $\Gamma$-limit.  Based on the 1-D CMFGEN models we conclude that, ultimately, the fast rotation of AG Car requires  $M \geq 60~\msun$ in order to satisfy the Eddington limit modified by rotation. It would be desirable to confirm this result with appropriate full 2-D, line-blanketed, non-LTE radiative transfer models when these become available.

\subsection{The evolutionary mass budget and current mass of AG Car: a massive star close to the Eddington limit}

The first constraint to the current mass of AG Car comes from evolutionary models. According to the non-rotating evolutionary models of \citetalias{meynet03}, and assuming an initial mass of $100~\msun$ and the derived surface He abundance (Y=0.62, \citetalias{ghd09}), the current mass of AG Car would be roughly $60~\msun$. This value is extremely close to the minimum mass of AG Car according to the modified Eddington limit derived above, which would imply that AG Car is {\it at} the $\Omega \Gamma$ limit. 

There are two important caveats of doing such a comparison. First, the evolutionary models of \citetalias{meynet03} do not include giant eruptions, while AG Car is known to have had a strong mass ejection $\sim10^4$ years ago.
Second, the evolutionary models of \citetalias{meynet03} include the mass-loss recipe during the Main Sequence (MS) and Blue Supergiant (BSG) phases from \citep{vink01}, while recent findings  (e.g., \citealt{bouret03}; see Sect. \ref{intro}) seem to suggest a lower mass-loss rate during these epochs. If that is true, the evolutionary models would predict that AG Car would reach the LBV phase with a lower $M$.

A second constraint to the current $M$ can be obtained by computing the mass budget throughout the evolution of AG Car. If we take into account that AG Car has a massive nebula ($15-30~\msun$) composed of material from the star \citep{lamers98} and an initial mass of $100~\msun$, then the mass budget is tight: AG Car would have $M \sim70-85~\msun$ assuming zero mass loss during the MS and BSG phases. A more realistic estimate of $\mdot\sim 2-4 \times 10^{-6}~\msunyr$ during $\sim3\times 10^6$ years would lead to $M \sim 60-65~\msun$. If the mass-loss rate during the MS and BSG phases was even lower, we could make the case for AG Car having $M\simeq 70~\msun$. {\it Therefore, evolutionary considerations and the presence of a massive nebula suggest that AG Car has $M\simeq60-70~\msun$ and is, thus, dangerously close to the Eddington limit modified by rotation}.

Obviously, a precise determination of the amount of nebular mass around AG Car and detailed evolutionary models that include a lower mass-loss rate during the MS than the one currently used, and including the mass lost in brief giant eruptions, are highly desirable.

\section{On the {\it modus operandi} of the S-Dor cycle toward maximum} \label{modus}

\begin{figure}
\resizebox{0.8\hsize}{!}{\includegraphics{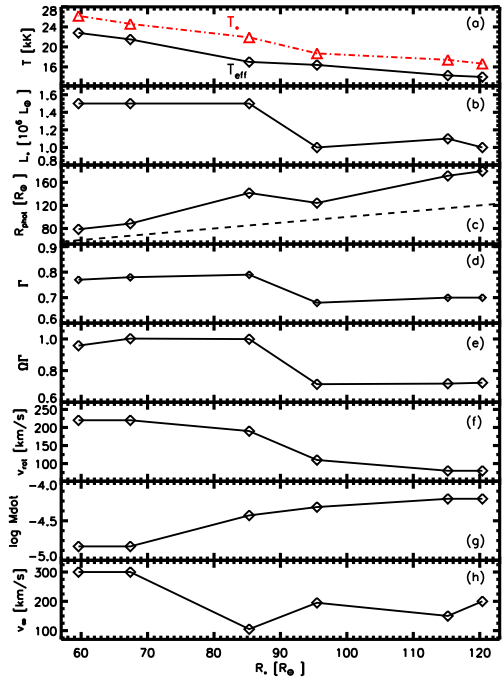}} 
\caption{\label{physparmin} Stellar and wind parameters of AG Car \citepalias{ghd09} as a function of the stellar radius obtained in different epochs as the star moves toward maximum. From top to bottom, we show the evolution of {\it (a)} stellar temperature at the base of the wind ($\tstar$) and effective temperature ($\teff)$, {\it (b)} bolometric luminosity ($\lstar$), {\it (c)} photospheric radius ($\reff$), {\it (d)} Eddington parameter ($\Gamma$), {\it (e)} Eddington parameter modified by rotation ($\Omega\Gamma$), {\it (f)} projected rotational velocity ($\vrot$), {\it (g)} mass-loss rate in units of $10^{-5}~\msunyr$ ($\mdot$), and {\it (h)} wind terminal velocity ($\vinf$).}
\end{figure}

The evolution of LBVs during their S-Dor cycles is commonly assumed to be due to changes in the stellar radius at constant luminosity \citep{leitherer89,humphreys89,leitherer94,sc94,shore96,dekoter96,stahl01,vink02}, which imply changes in the effective temperature. Based on a detailed spectroscopic analysis of multi-epoch observations of AG Car using CMFGEN, we argue that this scenario needs some revisions in order to explain the temporal behavior of the physical parameters of AG Car \citepalias{ghd09} which are shown in Fig. \ref{physparmin}.

We found that the stellar radius of AG Car changes during the S-Dor cycles, increasing from visual minimum to maximum, as has been determined by previous works (e.g \citealt{stahl01}). However, unlike previous works, using CMFGEN we can precisely derive both the hydrostatic radius of AG Car (\rstar, defined as $v=v_\mathrm{sonic}/3$) and the photospheric radius (\reff, defined at $\tau_\mathrm{Ross}=2/3$) for a given epoch, and thus, compare how they change as a function of time. We found that both $\rstar$ {\it and} $\reff$ increase from visual minimum to maximum, ruling out that the S-Dor type variability is caused only by an expanding pseudo-photosphere due to an increase in $\mdot$ \citep{leitherer85,lamers87,davidson87}. The increase in $\mdot$ is not high enough to explain the changes in $\reff$; definitely, our results show that a physical expansion of $\rstar$ is needed to explain the S-Dor type variability from visual minimum to maximum in the lightcurve, as first suggested by \citet{vg82}. The relatively high $\mdot$ derived for AG Car compared to B supergiants in the same temperature range, in combination with the proximity of the star to the Eddington limit, causes the unavoidable formation of an extended pseudo-photosphere, which translates into different values for $\rstar$ and $\reff$.

A major result from \citetalias{ghd09} is that the bolometric luminosity of AG Car decreases during the expansion of the hydrostatic radius, which is contrary to the aforementioned {\it status quo} explanation for the S-Dor variability. The detection of a decrease in $\lstar$ is not unique to AG Car; \citet{lamers95b} also detected that another famous, highly-variable LBV, S Doradus, presents a similar behavior. We support the hypothesis by \citet{lamers95b} that the decrease in $\lstar$ is due to the energy needed to expand the outer layers of the star. Variations in $\lstar$ were also detected in the LBV AFGL 2298 \citep{clark09}. However, in this latter case, the variation of $\lstar$ were apparently not related to changes in $\rstar$.

Another striking result is that the Eddington parameter $\Gamma$ of AG Car {\it decreases} as the star expands and its $\lstar$ and $\teff$ decrease (Fig. \ref{physparmin}). Such a reduction is, at least in part, caused by the decrease in $\lstar$. For AG Car, $\Gamma \simeq 0.78 $ during the hottest epochs, and decreases to $\Gamma \simeq 0.70 $ as the star evolves toward maximum. 

The Eddington parameter modified by rotation $\Omega\Gamma$ is even more dramatically reduced from visual minimum to maximum (Fig. \ref{physparmin}), since rotation will be relatively more important when the star has a higher $\teff$ (see discussion in \citealt{ghd06}). While AG Car is significantly close to the Eddington limit modified by rotation ($\Omega\Gamma \simeq 0.93 $ during 1985--1990 and 2001 April) during visual minimum, we found that $\Omega\Gamma \simeq 0.72 $ as the star expands. Therefore, if the proximity to the Eddington limit reflects the instability of the outer layers of AG Car, the star is more unstable during visual minimum than during maximum. 

The evolution of the wind parameters of AG Car is regulated by the bistability mechanism (Sect. \ref{compbi}). However, the behavior of the mass-loss rates and wind terminal velocities obtained in \citetalias{ghd09} do not closely follow the predictions of \citet{vink02}. This is due to several reasons: our modeling determined that AG Car is twice as massive as the value derived by \citet{vink02}, more luminous, and closer to the Eddington limit. Furthermore, AG Car has a high rotational velocity which varies from visual minimum to maximum \citep{ghd06}. Since the models from \citet{vink02} do not include rotation, significant differences are anticipated.  

\section{Conclusions } \label{conclusions}

The detailed spectroscopic analysis of AG Car using the radiative transfer code CMFGEN has provided a wealth of information on the LBV phenomenon and on the S-Dor type variability. Below, we summarize the main conclusions of this paper.

1. The high rotational velocity of AG Car detected by \citet{ghd06} during the visual minimum of 2001--2003 was also obtained during the previous visual minimum of 1985--1990, amounting to $\vrot\, {\mathrm sin\,i} \simeq 220 \kms$. We found that the rotational velocity is proportional to $\rstar^{-1.34 \pm 0.09}$, but including systematical errors which might arise due to the star being extremely close to the Eddington limit, we cannot rule out that $\vrot \propto \rstar^{-1}$.

2. We noticed that the rotational period of AG Car during the visual minimum of 1985--1990 ($\mathrm{P_{rot}}=13 \pm 2$ days) is remarkably similar to the cyclical photometric micro-variability of 11--14 days observed during these visual  minimum epochs \citep{vg88}. We suggest that the micro-variability period of AG Car might be modulated by the stellar rotation. If rotation is also responsible for the micro-variability seen in other LBVs, in principle, their rotational velocities can be derived using this method.

3. We found in \citetalias{ghd09} that the consecutive visual  minimum phases of 1985--1990
and 2000--2003 are different in duration, maximum magnitude achieved, maximum stellar temperature, mass-loss rate, and wind terminal
velocity. We suggest that these differences arise due to different stellar parameters, which cause different mass-loss rates and wind terminal velocities. We explain this behavior in terms of the bistability mechanism of line-driven winds, which is ultimately regulated by the optical depth of the Lyman continuum. From 1985--1990, AG Car was on the hot side of the bistability (\teff=22800 K), while the star was on the cool side (\teff=17000 K) from 2000--2003. Therefore, AG Car is the first star ever confirmed to be on different sides of the bistability at different epochs based on detailed spectroscopic analysis.

4. We noticed that AG Car has never been observed exactly {\it at} the boundary of the bistability limit, suggesting that significant instability is present when that happens. Although we identify AG Car in two regimes, the absence of observations close to the boundary of crossing from one side to the other of the bistability does not allow us to confirm the presence of a $jump$ in the wind properties of AG Car. An increase in $\mdot$ as a function of $\teff$ was found as the star crosses the bistability limit, while $\vinf$ decreases and presents a more complex behavior. 

5. The duration of the last two visual minimum phases of AG Car and the variability inside the visual minimum are related to the maximum
temperature achieved at each minimum and, therefore, on which side of the bistability the star was. It seems that visual  minimum phases that have 
$\teff$ high enough to put the star on the hot side of the bistability limit have a longer duration, 
while visual minimum phases which have $\teff$ low enough to put the star on the cool side, are brief. This behavior needs to be verified by further spectroscopic and photometric monitoring of AG Car during the next decades.

6. We suggest that the proximity to the Eddington limit modified by rotation ($\Omega\Gamma$) plays a key role in explaining the behavior of AG Car. The high value obtained for its luminosity ($\lstar=1.5\times10^{6} \lsun$), together with the high rotational velocity of AG Car detected by \citet{ghd06} and by this work, put the star extremely close to the Eddington limit modified by rotation, $\Omega\Gamma$ \citep{mm_omega00}. For an assumed mass of 70 $\msun$, $\Omega\Gamma \simeq 0.93 $ during 1985--1990 and 2001, and decreases to $\sim 0.73$ as the star expands towards maximum.

7. Despite the extreme conditions found in AG Car due to its high luminosity and fast rotation, the mass-loss rate remains relatively modest and does not wildly increase due to the proximity to the Eddington limit. Surprisingly, the mass-loss rate is even quite modest when compared to Eta Carinae, which has $\mdot$ two orders of magnitudes higher than AG Car, but a luminosity that is only 3 times higher.

8. Due to the proximity of AG Car to the Eddington limit and the mass loss, an extended pseudo-photosphere is formed, hiding the hydrostatic surface of the star. Even if the mass-loss rate increases towards maximum, that is not sufficient to explain the changes in the spectrum. An expansion in the hydrostatic radius is required in order to reduce the effective temperature.

9. Contrary to the current paradigm, the maximum value of the Eddington parameter is found during visual  minimum, not when the star is evolving towards maximum. Therefore, the picture of a quiescent star at visual minimum followed by an eruption during visual  maximum due to an increase in the Eddington parameter might not correctly describe the S-Dor cycles.

\acknowledgments

We thank the anonymous referee for a careful reading of the paper and useful comments. We are grateful to Nolan Walborn for making the 1989 March dataset available. We acknowledge discussions with Jorick Vink, Stan Owocki, Georges Meynet, Cyril Georgy, Nathan Smith, and Andre Maeder. We are thankful to Tom Madura and Jorick Vink for comments on the original manuscript. JHG thanks the Max-Planck-Gesellschaft (MPG) for financial support for this work. AD also acknowledge partial financial support from the Brazilian Agencies FAPESP and CNPq. DJH gratefully acknowledges partial support for this work from NASA-LTSA grant NAG5-8211. This research made use of the Smithsonian NASA/ADS and SIMBAD (CDS/Strasbourg) databases.

\end{document}